\documentstyle[12pt]{article}
\begin{document}
\rightline{TMUP-HEL-9207}
\rightline{November 1992}
\baselineskip=18pt
\vskip 0.7in
\begin{center}
{\large{\bf A TEST OF THE EQUIVALENCE PRINCIPLE BY}}
{\large{\bf LONG-BASELINE NEUTRINO-OSCILLATION EXPERIMENTS}}
\end{center}
\vskip 0.4in
\begin{center}
Kazuhito Iida, Hisakazu Minakata and Osamu Yasuda
\vskip 0.2in
{\it Department of Physics, Tokyo Metropolitan University}

{\it 1-1 Minami-Osawa Hachioji, Tokyo 192-03, Japan}
\end{center}

\vskip .7in
\centerline{ {\bf Abstract} }

We show that a breakdown of the universality of the gravitational
couplings to different neutrino flavors can be tested in long-baseline
neutrino-oscillation experiments.  In particular we have analyzed in
detail a proposed experiment at SOUDAN 2 with $\nu_\mu$ beams from the
Fermilab Main Injector.  It turns out that we can study both masses of
neutrinos and such a breakdown with sensitivity to the order of $10^{-14}$
by investigating the energy spectrum of the resulting muons.

\newpage

It has been pointed out${}^{1,2}$ that a breakdown of the
universality of the gravitational couplings to different neutrino flavors
could lead to neutrino oscillations.  In particular the authors of
Ref. 2 studied the possibility in which the solar neutrinos (see,
e.g., Ref. 3) can be used to test this kind of breakdown of the
universality.  Since the flux of the solar neutrinos is relatively small
and the energy spectrum is beyond our control, the utility of the solar
neutrino for this purpose is limited.  In this paper we propose a possible
long-baseline experiments of neutrino oscillations to test the breakdown of
the universality of the gravitational couplings to neutrinos.  In case of the
long-baseline experiments with an accelerator, we have a larger flux than that
of solar neutrinos and in principle we can change the energy spectrum of the
neutrino beams, and therefore more information, if any, can be obtained on
neutrino oscillations.  As we will see, the breakdown of the universality of
the gravitational couplings to neutrinos of different flavors leads to a
violation of Einstein's equivalence principle (see, e.g., Ref. 4)
which states that all the laws of physics must take on their familiar
special-relativistic forms in any and every local Lorentz frame, anywhere and
any time in the universe.  In our case, we show that we can probe the
magnitude of the breakdown of Einstein's equivalence principle to the order
of $10^{-14}$, assuming that there are neutrino mixings.  Among various
experiments to test the equivalence principle (see e.g., Ref. 5
for a review), there have been few tests of Einstein's equivalence principle
for neutrinos${}^{6}$.  The universality of the gravitational couplings
that we study in this paper is of different type from these experiments in
the past, so our discussions here are complementary to them.

In this paper we assume that there are two neutrino flavors which have
different couplings to gravity and that the eigenstates of these different
gravitational couplings do not coincide with those of the electroweak
flavors.  Throughout the present discussions we consider neutrino oscillations
between two flavors for simplicity.  Let us start with the following Lagrangian
of two kinds of neutrinos
\begin{eqnarray}
{\cal L}=&~&e(G_1){\overline \nu_1}\left[ ie^{a\mu}(G_1)
\gamma_a D_\mu(G_1)-m_1\right]\nu_1\nonumber\\
&+&e(G_2){\overline \nu_2}\left[ ie^{a\mu}(G_2)
\gamma_a D_\mu(G_2)-m_2\right]\nu_2\nonumber\\
&+& {\rm (interaction~ terms~ with~ electroweak~ gauge~ bosons)},
\label{eqn:Lagrangian}
\end{eqnarray}
where we have included mass terms to keep generality,
$e^{a\mu}(G_i)~(i=1,2)$ are the vierbein fields of some background metric
with different Newton constants $G_i~(i=1,2)$, and $e(G_i)\equiv\det
e^a_\mu(G_i)$.  For simplicity we assume that the eigenstates of the
gravitational couplings coincide with those of the masses.  Notice that
even if these neutrinos are massless, we cannot rotate these two fields
so that these are the eigenstates of the electroweak theory, since
the gravitational coupling terms are not invariant under the rotation
in the flavor space.  Note also that each term in (\ref{eqn:Lagrangian}) is
consistent with local Lorentz invariance, general covariance, and the CPT
invariance, as the gravitational couplings to particles and anti-particles
are the same.  Since the gravitational coupings for these two kinds of
neutrinos are different, even if we choose a coordinate system in which
the Dirac equation for $\nu_1$ in (\ref{eqn:Lagrangian}) becomes the one
in a flat space-time, the Dirac equation for $\nu_2$ in the same coordinate
system does not necessarily do so.  Thus Einstein's equivalence principle
is violated in (\ref{eqn:Lagrangian}).  The situation here is similar to that
of the gauge theory where the gauge invariance is explicitly broken, and
physics does depend on which gauge we choose.  So we are forced to choose one
particular coordinate system from which we start. The most natural choice in
our case seems to be the coordinate system which is at rest on the Earth.
This is because neutrinos which we observe are created from the accelerator and
annihilated near the detector, and both equipments are fixed on the Earth.
So we take the coordinate system which are moving together with the Earth,
and we choose as our background the so-called interior Schwarzschild metric
(see, e.g., chapter 11.7 in Ref. 7) whose curvature is entirely
caused by the gravitational field due to the Earth.  We will discuss the
issue of the choice of the coordinate systems again later.

The configuration of the long-baseline experiment we will discuss is depicted
in Fig. 1, and the neutrino beams go underneath the ground along the geodesics.
First let us consider the Dirac equation of left-handed neutrinos without any
flavor in the interior Schwarzschild background:

\begin{eqnarray}
 (ie^{a\mu}\gamma_a D_{\mu} - m)\psi = 0,
\end{eqnarray}
where $e_{a\mu}$ is the vierbein of the interior Schwarzschild metric
\begin{eqnarray}
ds^2 = (e^0_t)^2 dt^2 - (e^1_r)^2 dr^2 - (e^2_\theta)^2 d\theta^2
- (e^3_\varphi)^2 d\varphi^2
\label{eqn:metric}
\end{eqnarray}
and is given by
\begin{eqnarray}
e^0_t = {3 \over 2}\sqrt{1-{\alpha \over R}} - {1 \over 2}
\sqrt{1-{\alpha r^2 \over R^3}},~~e^1_r = {1 \over
\sqrt{1-{\alpha r^2 \over R^3}}},
{}~~e^2_\theta = r,~~e^3_\varphi = r\sin\theta.
\label{eqn:vierbein}
\end{eqnarray}
$D_\mu\psi\equiv(\partial_\mu -{1 \over 2}\omega_{\mu ab}\sigma^{ab})\psi$
is the covariant derivative acting on a spinor $\psi$, $\omega_{\mu ab}$
is the spin connection given by $e^b_{[\nu}\omega^a_{\mu ]b} = \partial_{[\mu}
e^a_{\nu]}$, and $\alpha$ in (\ref{eqn:vierbein}) is the Schwarzschild
radius.

One typical dimensionless parameter in our case is $ER$,
where $E$ is the energy of the neutrino, and $R$ is the radius of the
Earth.  For $E$=10 GeV and $R$=6,400 Km, $ER\sim 3\times10^{23}$, and
derivative terms in the spin connections are all of the order of $1/ER$,
so we will neglect them throughout this paper.  In this approximation
the Dirac equation becomes
\begin{eqnarray}
\left[ i(e^0_t)^{-1}\gamma^0{\partial{~} \over \partial t} + i(e^1_r)^{-1}
\gamma^1{\partial{~} \over \partial r} + i(e^2_\theta)^{-1}\gamma^2
{\partial{~} \over \partial \theta} + i(e^3_\varphi)^{-1}\gamma^3
{\partial{~} \over \partial \varphi} - m\right] \psi = 0
\label{eqn:Diracorig}
\end{eqnarray}
Since we consider neutrinos in the ultra relativistic limit $E\gg m$,
we have only to discuss (\ref{eqn:Diracorig}) along the geodesics for
massless fields.  The geodesics of the interior Schwarzschild metric is
given by
\begin{eqnarray}
{R^2\cos^2\delta \over r^2}=\sin^2\varphi+{3\alpha \over 2R}(\cos^2\delta
-\sin^2\varphi)
\label{eqn:geodesics}
\end{eqnarray}
to the order of $\alpha/R$, where $\delta$ is half of the angle $\angle$AOB
in Fig. 1.  We solve (\ref{eqn:Diracorig}) on the plane $\theta = \pi/2$,
and we remove the time dependence by $\psi({\vec x},t)=e^{-iEt}\chi({\vec x})$.
On the geodesics (\ref{eqn:geodesics}), (\ref{eqn:Diracorig}) becomes
\begin{eqnarray}
\left[ (e^0_t)^{-1}E\gamma^0-m+{1 \over 1+w^2}\left( (e^1_r)^{-1}w\gamma^1
+\gamma^3\right) {i \over r}{d{~} \over d\varphi}\right]\chi=0,
\label{eqn:Dirac0}
\end{eqnarray}
where $w\equiv d\ln r(\varphi)/d\varphi$, $r=r(\varphi)$ is defined in
(\ref{eqn:geodesics}), and we have taken only the tangential component into
consideration.  Following the convention of the Dirac matrices by Bjorken
and Drell${}^{8}$, and denoting $\chi^T\equiv(\chi_1,\chi_2,\chi_3,\chi_4)$,
it is easy to show that (\ref{eqn:Dirac0}) can be rewritten as
\begin{eqnarray}
{1 \over ir}{d{~} \over d\varphi}
\left(
\begin{array}{cc}
\chi_1+i\chi_2 \\ \chi_3-i\chi_4
\end{array}
\right) = \sqrt{(e^0_t)^{-2}E^2-m^2}
\sqrt{1+w^2 \over 1+(e^1_r)^{-2}w^2}
{}~U\sigma_3 U^{-1}
\left(
\begin{array}{cc}
\chi_1+i\chi_2 \\ \chi_3-i\chi_4,
\end{array}
\right)
\end{eqnarray}
where U is a certain $2\times 2$ matrix.
It turns out that a derivative term $r^{-1}dU^{-1}/d\varphi$ is of order
$1/ER$ which is extremely small, and hence we get
\begin{eqnarray}
{1 \over ir}{d{~} \over d\varphi}
\left(
\begin{array}{cc}
\nu\\ \widetilde\nu
\end{array}
\right) = \sqrt{(e^0_t)^{-2}E^2-m^2}
\sqrt{1+w^2 \over 1+(e^1_r)^{-2}w^2}
{}~\sigma_3\left(
\begin{array}{cc}
\nu\\ \widetilde\nu
\end{array}
\right),
\label{eqn:Dirac1}
\end{eqnarray}
where $(\nu,\widetilde\nu)\equiv (\chi_1+i\chi_2,\chi_3-i\chi_4)
(U^{-1})^T$.
$\nu$ and $\widetilde\nu$ correspond to the forward-going and the
backward-going energy solutions, and we will consider only the
forward-going solution in the following.  $\alpha$ in eq. (\ref{eqn:vierbein})
is the Schwarzschild radius of the Earth which is about 9 mm, so we expand
(\ref{eqn:Dirac1}) to the first order in $\alpha/r$.  Thus, in the ultra
relativistic limit $E\gg m$, we obtain
\begin{eqnarray}
{1 \over i}{d\nu \over dx}=
B\left[ 1-{m^2 \over 2E^2}+{\alpha \over R}\left(
{3 \over 2}-\cos^2\delta-2x^2\cos^2\delta\right)
\right] \nu,
\end{eqnarray}
where we have defined a variable $x\equiv \tan(\varphi-\pi/2)$, and
$B\equiv ER\cos\delta$ is a very large number.
Now let us go back to the Lagrangian (\ref{eqn:Lagrangian}) with two
kinds of neutrinos.  This time we assume that the vierbein fields
in the Lagrangian (\ref{eqn:Lagrangian}) are those of the interior
Schwarzschild background (\ref{eqn:vierbein}).  From the previous discussion,
it is straightforward to see that the Dirac equation for
(\ref{eqn:Lagrangian}) is given by
\begin{equation}
{1 \over i}{d{~} \over dx}\left(
\begin{array}{cc}
\nu_1 \\ \nu_2
\end{array}
\right)
=B\left[ 1-{m^2_1+m^2_2 \over 4E^2}+
(f_1+f_2)\Phi\left( {3\over 2}-\cos^2\delta-2x^2\cos^2\delta\right)
+\Delta(x)
\sigma_3\right] \left(
\begin{array}{cc}
\nu_1 \\ \nu_2
\end{array}
\right),
\label{eqn:Dirac2}
\end{equation}
where
\begin{eqnarray}
\Delta(x)\equiv {\Delta m^2 \over 4E^2}+\Delta f\Phi\left(
{3 \over 2}-\cos^2\delta-2x^2\cos^2\delta\right) .
\end{eqnarray}
Here $\Delta m^2\equiv m^2_2-m^2_1$ is the
difference of the masses, we have defined the Newton potential
$\Phi\equiv -GM/R$ on the surface of the Earth, and we have also defined
the difference $\Delta f=f_2-f_1$ of the dimensionless gravitational
couplings of the two neutrino species
\begin{eqnarray}
\left(
\begin{array}{cc}
f_1\Phi \\ f_2\Phi
\end{array}
\right)
=-\left(
\begin{array}{cc}
{\alpha_1/2R} \\ {\alpha_2/2R}
\end{array}
\right)
=-\left(
\begin{array}{cc}
{G_1M/R} \\ {G_2M/R}
\end{array}
\right) .
\end{eqnarray}
The equation (\ref{eqn:Dirac2}) can be easily integrated from $x=-\tan\delta$
to $x=\tan\delta$.

Now let us introduce the flavor eigenstates $\nu_a,~\nu_b$ of the weak
interaction by
\begin{eqnarray}
\left(
\begin{array}{cc}
\nu_a \\ \nu_b
\end{array}
\right)
=
\left(
\begin{array}{cc}
\cos\theta&-\sin\theta\\
\sin\theta&\cos\theta
\end{array}
\right)
\left(
\begin{array}{cc}
\nu_1\\ \nu_2
\end{array}
\right) .
\end{eqnarray}
Then the probability of detecting a different flavor $\nu_b$ at a distance
$L$ after producing one neutrino flavor $\nu_a$ is given by
\begin{eqnarray}
P(\nu_a\rightarrow\nu_b)=\sin^22\theta~\sin^2\left[
\left( {\Delta m^2 \over 4E^2} + {\Delta f\Phi \over 2}\left(
1+{L^2 \over 6R^2}\right)
\right) EL\right] ,
\label{eqn:prob1}
\end{eqnarray}
where $L\equiv 2R\sin\delta$ is the distance AB in Fig. 1.
This formula applies to the transition between $\nu_\mu$ and $\nu_\tau$,
where no MSW effect${}^{9}$ is expected to occur.

In case of the transition between $\nu_e$ and $\nu_\mu$, we have to take
the MSW effect${}^{9}$ into consideration, and the Dirac equation
is modified as
\begin{eqnarray}
{1 \over i}{d{~} \over dx}\left(
\begin{array}{cc}
\nu_e\\ \nu_\mu
\end{array}
\right) =B\left(
\begin{array}{cc}
\Delta(x)\cos 2\theta-{G_FN_e \over \sqrt{2}E}&\Delta(x)\sin 2\theta\\
\Delta(x)\sin 2\theta&-\Delta(x)\cos 2\theta+{G_FN_e \over \sqrt{2}E}
\end{array}
\right) \left(
\begin{array}{cc}
\nu_e\\ \nu_\mu
\end{array}
\right) ,
\label{eqn:Dirac3}
\end{eqnarray}
where $G_F$ is the Fermi coupling constant, and $N_e$ is the density of
electrons in the Earth.  (\ref{eqn:Dirac3}) can be solved in the same way
as before by introducing the variables
\begin{eqnarray}
\Delta_N(x)\cos 2\theta_N&=&\Delta(x)\cos 2\theta-{G_FN_e \over \sqrt{2}E}
\nonumber\\
\Delta_N(x)\sin 2\theta_N&=&\Delta(x)\sin 2\theta.
\label{eqn:DeltaN}
\end{eqnarray}
Note that $\theta_N$ does depend on the variable $x$ in this case.
It is easy to integrate (\ref{eqn:DeltaN}), and we have the transition
probability of detecting $\nu_e$ at a distance $L$ from the source of
$\nu_\mu$ beams
\begin{eqnarray}
P(\nu_\mu\rightarrow\nu_e)=\sin^2 2\theta_N( x=\tan\delta)
\sin^2\left( \int^{\tan\delta}_{-\tan\delta}dx\,\Delta_N(x)\right),
\label{eqn:prob2}
\end{eqnarray}
where we have used the fact $\theta_N(x=\tan\delta)=\theta_N(x=-\tan\delta)$
and $\Delta_N$, $\theta_N$ are defined through (\ref{eqn:DeltaN}).
We have performed the integration in the exponent in (\ref{eqn:prob2})
numerically.

Here we would like to comment again on the dependence of our results on
the choice of the coordinate systems.  As we mentioned earlier, equations
(\ref{eqn:Dirac1}) and (\ref{eqn:Dirac2}) depend on how we choose
a coordinate system, since the Newton potential terms in (\ref{eqn:Dirac1})
and (\ref{eqn:Dirac2}) are changed if we switch to a coordinate system
which moves with acceleration relative to the Earth.  In Ref. 10
it was argued that one could derive stronger bound on a breakdown of the
equivalence principle by using the contribution to the Newton potential
from the supergalactic cluster.  However, this argument holds only when
one assumes the Lagrangian (\ref{eqn:Lagrangian}) in the coordinate system
which is at rest in the supergalactic cluster.  Such a coordinate system
is different from our coordinate system, i.e., a different choice of gauge
is taken in Ref. 10.  Throughout this paper we take the
background (\ref{eqn:vierbein}), and we make our analyses below using
this choice of the background.  Since any other contribution to the Newton
potential is negative definite, the absolute value of $\Phi$ in
(\ref{eqn:prob1}) with our ansatz is the smallest among all possibilities.
Hence our choice gives the most conservative bound on $\Delta f$.

Let us now consider a proposed long-baseline neutrino-oscillation
experiment which will be performed with $\nu_\mu$ beams from the Fermilab Main
Injector${}^{11}$.  Our discussions here are analogous to those by
Bernstein and Park${}^{12}$.  Although there may be several factors
which cause the systematic errors as has been emphasized in Ref. 12,
we will not discuss this issue in this paper.  We assume that the Fermilab
Main Injector neutrino beams have an energy spectrum which is given by
Fig. 6.27 in Ref. 11, but we extrapolate the graph in
Ref. 11 naively for neutrinos of the energy larger than 70 GeV.
The average of the energy of the neutrino beams is typically from 10 GeV to
20 GeV.  If Einstein's equivalence principle is violated, the higher the
energy of the neutrino beams becomes, the more probability of neutrino
oscillations we have (see eq. (\ref{eqn:prob1})), unlike in case of
neutrino-oscillations due to masses.  We choose a value of the distance
$L$=800 Km which is motivated by the SOUDAN 2 experiment.  We mainly consider
the so-called disappearance experiments${}^{12,11}$, in
which the initial $\nu_\mu$ flux at some short distance and the $\nu_\mu$ flux
at the detector are measured by detecting muons which are created from
charged-current interactions.  The probability of detecting muons at a
distance $L$ is given by $1-\epsilon(\nu_\mu\rightarrow\nu_a)~(a=e~{\rm or}
{}~\tau)$ where
\begin{eqnarray}
\epsilon(\nu_\mu\rightarrow\nu_a)\equiv
{\int^{q_{\rm max}}_0\epsilon (q) dq\int^{E_{\max (q)}}_{E_{{\rm min} (q)}}
{d\sigma (E,q) \over dq} F(E) P(\nu_\mu\rightarrow\nu_a) n_T(q)dE
\over
\int^{q_{\rm max}}_0\epsilon (q) dq\int^{E_{\max (q)}}_{E_{{\rm min} (q)}}
{d\sigma (E,q) \over dq} F(E) n_T(q)dE}~~(a=e~{\rm or}~\tau).
\label{eqn:epsilon}
\end{eqnarray}
Here $E$ and $q$ are the energy of the incident $\nu_\mu$ and the outgoing
muon, respectively, $F(E)$ is the flux of neutrino beams, $n_T(q)$ is the
effective number of target nucleons, we have
modeled the detection efficiency function $\epsilon (q)$ for
muons with a step function, as in Ref. 12, and we have used
the $y$ distribution for deep-inelastic scattering of neutrinos to determine
the cross section $d\sigma(E,q)/dq$ of charged-current interaction${}^{13}$.
$P(\nu_\mu\rightarrow\nu_e~{\rm or}~\nu_\tau)$ in
(\ref{eqn:epsilon}) is either $P(\nu_\mu\rightarrow\nu_e)$ or $P(\nu_\mu
\rightarrow\nu_\tau)$, depending on whether we consider oscillations
$\nu_\mu\leftrightarrow\nu_e$ or $\nu_\mu\leftrightarrow\nu_\tau$.

We have studied the quantity $\epsilon$ for various cases.  In case of
$\nu_\mu\leftrightarrow\nu_\tau$ oscillations, there is no MSW effect,
and we have obtained the results for $\epsilon(\nu_\mu\rightarrow\nu_\tau)$
from (\ref{eqn:prob1}) and (\ref{eqn:epsilon}).  We have plotted,
assuming no signal for beam attenuation up to $\epsilon$, the
excluded region of the $(\sin^2 2\theta, \Delta f/10^{-14})$ plane for
$\nu_\mu\leftrightarrow\nu_\tau$ oscillations with the detection threshold
energy $E_{\rm th}$=10 GeV of muons in Fig. 2.  We have also calculated the
excluded region for $E_{\rm th}$=5 and 20 GeV, and the results are almost
similar to the case of $E_{\rm th}$=10 GeV.  In Fig. 3 we have plotted the
excluded region of the $(\sin^2 2\theta, -\Delta f/10^{-14})$ plane for
$\nu_\mu\leftrightarrow \nu_e$ with $E_{\rm th}$=10 GeV.  To get this region
we have used (\ref{eqn:prob2}) and (\ref{eqn:epsilon}), where the MSW effect
is taken into account.  We have used a value for the density of electrons:
$N_e=8.05\times10^{23}$ electrons/cm$^3$ which is constant
on the entire trajectory of neutrino beams in the present case (see, e.g.,
Ref. 14).   Comparing Fig. 2 with Fig. 3, we observe that there is a
slight difference in $\nu_\mu\leftrightarrow\nu_e$ oscillations because
of the MSW effect.  We have shown the result with negative $\Delta f$ in
Fig. 3, because (\ref{eqn:DeltaN}) shows that negative $\Delta f$ gives larger
$\vert\Delta_N(x)\vert$ and therefore $\epsilon (\nu_\mu\rightarrow\nu_e)$
with negative $\Delta f$ is larger than that with the same $\Delta m^2$ and
positive $\Delta f$.  Notice that $\epsilon (\nu_\mu\rightarrow\nu_\tau)$
with $\Delta m^2=0$ and $-\Delta f$ is the same as $\epsilon (\nu_\mu
\rightarrow\nu_\tau)$ with $\Delta m^2=0$ and $\Delta f$, as is obvious
from (\ref{eqn:prob1}).

One important feature is the dependence of $\epsilon$ on the detection
threshold energy $E_{\rm th}$ of muons.  As we mentioned earlier, it is more
advantageous to look at neutrinos (and therefore outgoing muons) of higher
energy to investigate a breakdown $\Delta f$ of the universality of the
gravitational couplings.  We have shown $\epsilon(\nu_\mu\rightarrow\nu_e)$
as a function of $E_{\rm th}$ in Figs. 4 and 5 with $\sin^22\theta=0.5$
and with $\Delta m^2=1\times 10^{-2}\,{\rm eV}^2,0.5\times 10^{-2}\,{\rm
eV}^{-2}$, and 0.  The parameters are $\Delta f=0$ for Fig. 4 and $\Delta
f=-0.5\times 10^{-14}$ for Fig. 5, respectively.  $\epsilon(\nu_\mu\rightarrow
\nu_e)$ with $\Delta f=0$ in Fig. 4 decreases as $E_{\rm th}$ increases, since
the argument of the phase (=the mass squared) in case of ordinary MSW effect
(see (\ref{eqn:prob2})) is suppressed by the neutrino energy $E_\nu$ $(\propto
1/E_\nu)$.  Fig. 5 shows, on the other hand, that $\epsilon(\nu_\mu\rightarrow
\nu_e)$ in the presence of $\Delta f(=-0.5\times 10^{-14})$ has a conspicuous
difference from the case with $\Delta f=0$.  Thus, if we look at the $E_{\rm
th}$-dependence of $\epsilon(\nu_\mu\rightarrow\nu_e)$, we can determine
$\Delta f$ with higher accuracy.  Because of the interaction term $G_FN_e$
with matter, $\epsilon(\nu_\mu\rightarrow\nu_e)$ is in general larger than
$\epsilon (\nu_\mu\rightarrow\nu_\tau)$, but the qualitative features for the
results of $\nu_\mu\leftrightarrow\nu_\tau$ oscillations are the same as those
of $\nu_\mu\leftrightarrow\nu_e$.  The dependence of $\epsilon$ on $E_{\rm th}$
is also very useful, if we can change the energy of the neutrino beams.
Namely, by comparing the results with different two values of energy
of the neutrino beams, we could establish the existence of non-zero value
of $\Delta f$.  If the statistics is good enough, then we could even measure
masses of neutrinos, by subtracting the effects of $\Delta f\not=0$ from
the data.  This is one advantage that long-baseline experiments have
over solar neutrino experiments.

Another important point is the expected numbers of observed muons.
In case of the planned experiments at SOUDAN 2, we estimate
the numbers to be approximately 16,000 events/year for $E_{\rm th}=10$ GeV
and 6,000 events/year for $E_{\rm th}=20$ GeV, which are significantly larger
than those of solar neutrino experiments where typical numbers are several
hundreds events/year.  So also in this aspect long-baseline experiments are
promising.

In this paper we have proposed long-baseline experiments to test the
universality of the gravitational couplings of neutrinos, and we found
that we could probe the dimensionless parameter $\Delta f$ as small as
$10^{-14}$ which is smaller by a few orders of magnitudes than the upper
limit on a breakdown of the equivalence principle from different types
of experiments.  Although we have not evaluated systematic errors in
detail, we hope our analysis will stimulate and motivate long-baseline
experiments in the near future.
\vskip 0.2in
\noindent
{\Large{\bf Noted Added}}
\vskip 0.1in

Toward the completion of our paper, we became aware of the work by
Pantaleone, Halprin and Leung${}^{15}$, where the similar
topics has been discussed from a slightly different viewpoint.
\vskip 0.2in
\noindent
{\Large{\bf Acknowledgement}}
\vskip 0.1in

We would like to thank C.N. Leung for discussions and informing us
of their work prior to publication.  We also would like to thank
members of the astrophysics group of our university for discussions.
We have benefited conversations with numbers of participants at the
International Symposium on Neutrino Astrophysics at Takayama, Japan
in October 1992.

\vskip 0.2in
\noindent
{\Large{\bf References}}
\begin{enumerate}
\item M. Gasperini, Phys. Rev. {\bf D38} (1988) 2635; {\it ibid}
{\bf D39} (1989) 3606.
\item A. Halprin and C.N. Leung, Phys. Rev. Lett. {\bf 67} (1991)
1833.
\item J. N. Bahcall, ``Neutrino Astrophysics'',
Cambridge University Press (1989).
\item C.W. Misner, K.S. Thorne and J.A. Wheeler,
``Gravitation'', W.H. Freeman and Company (1973).
\item C.M. Will, Int. J. Mod. Phys. {\bf 1} (1992) 13.
\item M.J. Longo, Phys. Rev. Lett. {\bf 60} (1987) 173; L.M. Krauss
and S. Tremaine, {\it ibid.} 176.
\item C. M\o ller, ``The Theory of Relativity'', 2nd ed.,
Oxford University Press (1972).
\item J.D. Bjorken and S.D. Drell, ``Relativistic Quantum
Mechanics'', McGraw-Hill, Inc. (1964).
\item S.P. Mikheyev, A. Yu. Smirnov,
Sov. J. Nucl. Phys. {\bf 42} (1985) 913;
L. Wolfenstein,  Phys. Rev. {\bf D17} (1987) 2369
\item I.R. Kenyon, Phys. Lett.  {\bf B237} 274 (1990).
\item R. Bernstein {\it et al.}, Conceptual Design Report:
Main Injector Neutrino Program, Fermilab, June 1991.
\item R.H. Bernstein and S.J. Parke, Phys. Rev. {\bf D44} (1991)
2069.
\item P.S. Auchincloss {\it et al.}, Z. Phys. {\bf C 48}
(1990) 411.
\item F.D. Stacey, ``Physics of the Earth'', 2nd ed.,
John Wiley \& Sons, Inc. (1977).
\item J. Pantaleone, A. Halprin and C.N. Leung, Univ. of
Delaware preprint, UDHEP-10-92 (1992).
\end{enumerate}

\vskip 0.2in
\noindent
{\Large{\bf Figures}}

\begin{enumerate}
\item The cross section of the Earth.  The accelerator is located at
point A, and the SOUDAN 2 detector is located at point B.  The distance
$L$ between the points A and B is about 800 Km, and the radius $R$ of
the Earth is about 6,400 Km.  The curve from A to B in the Earth is
geodesics in the background of the interior Schwarzschild metric.

\item The excluded region in the $(\sin^2 2\theta,\Delta f/10^{-14})$
plane for $\nu_\mu\leftrightarrow\nu_\tau$ with $\epsilon$=1\% (solid),
3\% (dashed) and 10\% (dotted line).  The detection threshold energy of
muons is 10 GeV.  The upper and right side of the curves is excluded.

\item The excluded region in the $(\sin^2 2\theta,-\Delta f/10^{-14})$
plane for $\nu_\mu\leftrightarrow\nu_e$ with $\epsilon$=1\% (solid),
3\% (dashed) and 10\% (dotted line).  The detection threshold energy of
muons is 10 GeV.  The upper and right side of the curves is excluded.

\item The transition probability $\epsilon(\nu_\mu\rightarrow\nu_e)$
as a function of the detection threshold energy $E_{\rm th}$ of muons for
$\sin^22\theta=0.5$ and $\Delta f=0$, which is solely due to the standard
MSW effect.  The solid and dashed curves have parameters $\Delta m^2=1\times
10^{-2} {\rm eV}^2$ and $0.5\times 10^{-2} {\rm eV}^2$, respectively.

\item The transition probability $\epsilon(\nu_\mu\rightarrow\nu_e)$
as a function of the detection threshold energy $E_{\rm th}$ of muons for
$\sin^22\theta=0.5$ and $\Delta f=-0.5\times 10^{-14}$.  The solid, dashed
and dotted curves have parameters $\Delta m^2=1\times 10^{-2} {\rm eV}^2$,
$0.5\times 10^{-2} {\rm eV}^2$, and 0, respectively.

\end{enumerate}

\end{document}